\documentstyle{elsart}
\input epsf

\begin{document}
\begin{frontmatter}
\title{High-energy exclusive leptoproduction \\
of vector mesons\thanksref{meet}
}
\thanks[meet]{to appear in the proceedings of the workshop ELFE at DESY}

\author{Thierry Gousset}

\address{
Institut f\"ur Theoretische Physik, Philosophenweg 19, D-69120 Heidelberg\\
NIKHEF, P.O. Box 41882, NL-1009 DB Amsterdam}

\begin{abstract}
The physics of diffractive vector meson production in virtual 
photon nucleon scattering at NMC energies is reviewed. A particular 
attention is paid to the physical aspects of the reaction and how 
they influence the observables. The reaction is a good probe to 
investigate both soft exchange mechanisms and hadronic wave functions. 
Extension to either HERA or ELFE kinematics is sketched out.
\end{abstract}

\end{frontmatter}

\section{Space-time aspect of the reaction}

\begin{figure}
$$
\epsfysize=4cm\epsfbox{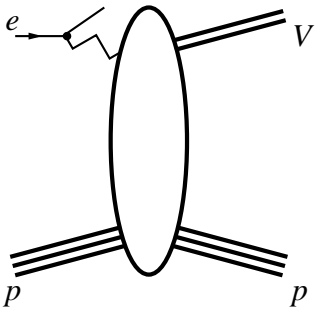}
$$
\caption{The reaction $\gamma^*+p\to V+p$. The initial photon 
4-momentum is denoted with $q$, the initial nucleon momentum with 
$p$, and the equivalent final states with $q'$ and $p'$.}
\label{gamma-p}
\end{figure}

The reaction of interest is, within the one-photon exchange approximation, 
\[\gamma^*+p\to V+p,\] where $V$ is a vector meson ($\rho$, $\phi$, 
$J/\psi$, etc.). It is depicted in Fig.~\ref{gamma-p}. The present review 
focuses on vector meson production off a proton target. The extension to 
nuclear targets is of great importance to observe the phenomenon of color 
transparency. It has also been discussed during the meeting~\cite{kop96}. 

We are interested in high-energy processes, i.e. when the center of mass 
energy $W=\smash{\sqrt{(q+p)^2}}$ is of the order of 10$\,$GeV or more, 
and in the low momentum transfer region, $-t=-(q-q')^2<1\,$GeV$^2$. 
In this kinematical domain the strong interaction is predominantly 
mediated by the pomeron. What are the precise properties of the 
pomeron and how such an object can emerge from QCD is still under 
debate, hence the interest of the field. In order to have a simple 
behavior of the photon, we will also limit ourselves to the region of 
large $Q^2=-q^2$, say $Q^2>2\,$GeV$^2$, but small $x_B=Q^2/2p.q<0.1$.

\begin{figure}[h]
$$
\epsfysize=3cm\epsfbox{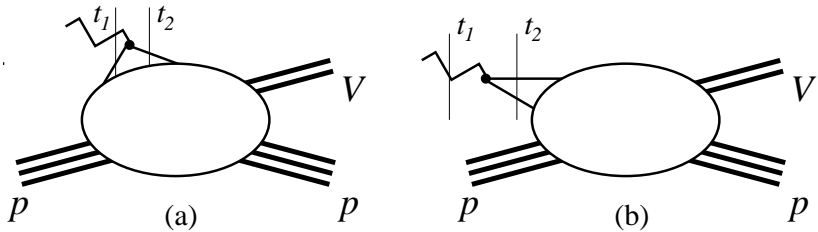}
$$
\caption{The two possible time-ordered configurations ($t_1<t_2$).}
\label{intermediate-state}
\end{figure}

The fact that reactions induced by a photon, either real or virtual, 
can help to get insights into the diffractive domain of strong 
interaction is due to the tendency of the photon to split up into 
partonic components~\cite{bau78}. To see this, let us consider, in 
the proton rest frame, a photon flying in the $z$-direction so that 
$q=(\nu,{\bf 0}_T,\sqrt{\nu^2+Q^2})$. The two possible time-ordered 
configurations are depicted in Fig.~\ref{intermediate-state}, where:
\begin{itemize}
\item (a) the photon directly attaches to a proton constituent;
\item (b) the photon fluctuates into a $q\bar{q}$-pair before 
reaching the proton.
\end{itemize}

The strength of each of these possibilities is a matter of energy 
denominators between initial and intermediate states. In 
Fig.~\ref{intermediate-state}(a), the quark energy is 
$E=(K_z^2+{\bf K}_T^2+m_q^2)^{1/2}$ before the photon vertex and 
$E'=((q_z+K_z)^2+{\bf K}_T^2+m_q^2)^{1/2}$ after. The magnitude of the 
initial quark momentum components, $K_z$ and ${\bf K}_T$, is fixed 
by the proton wavefunction and is $O(1/R_p)$, with $R_p$ the proton 
radius. This leads to the estimate $\Delta E_a=E+\nu-E'=O(1/R_p)$.
In Fig.~\ref{intermediate-state}(b), we have 
$E_q=(z^2(\nu^2+Q^2)+{\bf k}_T^2+m_q^2)^{1/2}$ for the quark, where 
$z=k_z/q_z$ is the longitudinal fraction it carries. To get the antiquark 
energy, $E_{\bar{q}}$, we replace $z$ by $1-z$ in $E_q$. The magnitude 
of ${\bf k}_T$ and $z$ is now determined by the vector meson wave 
function. This leads to \[
(\Delta E_b)^{-1}={2\nu\over Q^2+{{\bf k}_T^2+m_q^2\over z(1-z)}}
\approx(x_BM_p)^{-1},
\] which dominates $(\Delta E_a)^{-1}$ in the small $x_B$ region we 
are considering.

This can be reformulated by saying that the $q\bar{q}$-fluctuation has 
a lifetime or {\em coherence length}
$$
l_c=(E_q+E_{\bar{q}}-\nu)^{-1}\approx(x_BM_p)^{-1},
$$
much larger than the interaction time, i.e. the  proton radius, at 
very small $x_B$. As a consequence, the proton practically always sees 
the photon as a quark-gluon wave packet in this regime. In the proton 
rest frame (and in the center mass frame as well), a simple physical 
picture of the process emerges where the reaction is a three-step process:
\begin{itemize}
\item the photon turns into a quark-gluon state,
\item this state then scatters softly on the proton constituents,
\item before it forms the final vector meson.
\end{itemize}
From this physical evolution of the reaction, we deduce that 
experimental studies of vector meson electroproduction help us 
to learn about both hadronic wavefunctions and soft exchange 
mechanism (``pomeron exchange'').

To be more specific, the amplitude of the process can be written as 
\begin{equation}
\label{amplitude}
{\cal M}(\gamma^*+p\to V+p)=\int dz d^2{\bf r}\,\psi_{\gamma^*}(z,{\bf r})
{\cal M}(q\bar{q}+p\to q\bar{q}+p)\psi^*_V(z,{\bf r}),
\end{equation}
where $\psi_{\gamma^*}$ and $\psi_V$ are the photon and vector meson 
light-cone wave function, respectively. ${\cal M}(q\bar{q}+p)$ 
is the interaction amplitude between the $q\bar{q}$-dipole, with 
transverse extension $\bf r$, and the proton. It can be in turn expanded 
in terms of the initial and final proton wavefunctions and a 
fully microscopic interaction amplitude at the quark-gluon level.
The process is thus similar to the high-energy elastic collisions 
of a meson with a proton, in which case the amplitude reads
$$
{\cal M}(M+p\to M+p)=\int dz d^2{\bf r}\,\psi_M(z,{\bf r})
{\cal M}(q\bar{q}+p\to q\bar{q}+p)\psi^*_M(z,{\bf r}),
$$
but the photon virtuality $Q^2$ and polarization (either longitudinal 
or transverse) provide extra degrees of freedom in addition to the 
center of mass energy $\sqrt{s}\equiv W$ and momentum transfer $t$.

The phenomenology associated with the photon degrees of freedom is 
determined from its wavefunction. The simplest $q\bar{q}$-component 
can be easily computed in perturbation theory. Its main property 
is that large transverse distances are suppressed by a factor 
$e^{-\varepsilon r}$ where $\varepsilon=\sqrt{z(1-z)Q^2+m_q^2}$. 
In vector meson production, it is actually the wave function overlap, 
$\psi_{\gamma^*}\psi^*_V$,
which matters, so that the end-point region in $z$ is relatively 
unimportant, and the transverse distance probed diminishes with 
increasing $Q^2$. Furthermore, it turns out that this overlap depends 
on the photon polarization, the important differences being that 
\begin{equation}
\left.\psi_{\gamma^*}\psi^*_V\right|_L\propto z(1-z)Q,\ 
\left.\psi_{\gamma^*}\psi^*_V\right|_T\propto\varepsilon r,
\end{equation}
from which we can infer that the longitudinal transition dominates 
the transverse one at large $Q^2$. 
 
\section{The Donnachie-Landshoff Model~\protect\cite{don87}}

A phenomenological step towards the understanding of soft 
collisions can be accomplished by extracting their universal 
characteristics. This can be done by discussing the model of 
Donnachie and Landshoff. The model is based on a Regge approach 
and aims at describing every high-energy elastic collisions with 
a few ingredients. Despite (thanks to?) its simplicity, it gives 
a reasonable account of the present hadron-hadron elastic 
scattering data.

\begin{figure}[h]
$$
\epsfysize=3cm\epsfbox{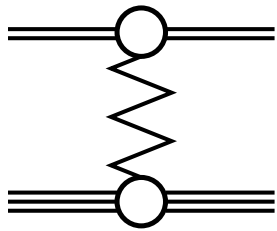}
$$
\caption{Diagram for $\pi$--$p$ elastic scattering.}
\label{pi-p}
\end{figure}

The model has a diagrammatic formulation and the relevant diagram 
for pion-proton collision at large center of mass energy, 
$\sqrt{s}$, and small momentum transfer, $t$, is shown in 
Fig.~\ref{pi-p}.

In Fig.~\ref{pi-p}, the exchanged object is a pomeron to which one 
associates the factor
\begin{equation}\label{pomeron}
2i\left({s\over s_0}\right)^{\alpha_0+\alpha' t},
\end{equation}
where the pomeron intercept, $\alpha_0$, is observed to be 
$\alpha_0\approx 1.08$ and the slope 
$\alpha'\approx 0.25\,$GeV$^{-2}$. Other objects (reggeons) can 
be exchanged in the reaction. Their contributions are however 
subleading in energy, i.e. suppressed by some powers of $1/s$, 
and will be ignored for simplicity.

The second necessary ingredient in Fig.~\ref{pi-p} is the 
hadron-pomeron vertex. Phenomenologically, one obtains a good 
description assuming that the pomeron couples to quarks like a photon. 
Of course, it does not interact with the 
individual quarks via their electromagnetic charge but rather 
couples to each {\em constituent} quark with a strength $\beta_0$. 
The pion-pomeron and proton-pomeron coupling are then \[
(\pi-P-\pi)=2\beta_0 F_{\pi}(t),\ \ \ \ \ (p-P-p)=3\beta_0 F_1(t),\]
where the $t$-dependence of the vertex is given by the hadron 
leading form factor.

As already stated, this model provides a good description of 
total cross sections which are related to the elastic amplitude 
through the optical theorem \[
\sigma(s)={1\over s}\,{\rm Im}\,{\cal M}(s,t=0),
\] with a universal ``slow'' rise $s^{0.08}$ in the $\sim 100\,$GeV 
region. Including the next-to-leading trajectory, the model 
provides good fits down to $\sqrt{s}=5\,$GeV for $p+p$, $\bar{p}+p$, 
$\pi+p$, $K+p$, $\gamma+p$~\cite{rpp96}.

The fact that the pomeron counts the number of constituents 
inside the hadron is called {\em quark additivity}. With this 
property, wave function effects expected from Eq.~(\ref{amplitude}) 
are hidden inside the form factor that enters into the 
hadron-pomeron coupling.

\begin{figure}[h]
$$
\epsfysize=3cm\epsfbox{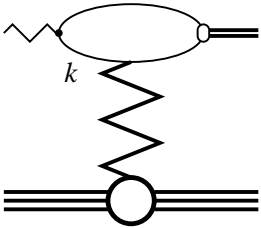}
$$
\caption{Diagram for exclusive $\rho$-electroproduction at large $Q^2$.}
\label{rho-prod}
\end{figure}

Let us now return to the exclusive electroproduction of 
$\rho$-mesons. In the Donnachie-Landshoff model, the relevant graph 
is depicted in Fig.~\ref{rho-prod}. 

The only difference with the diagram describing pion-proton 
scattering, Fig.~\ref{pi-p}, is that
the pion-pomeron vertex is now replaced by a quark-loop. Assuming 
that a non-relativistic description of the $\rho$-meson suffices 
to evaluate the quark-loop, one sees that the quark with momentum 
$k$ is far off-shell with $k^2\approx -Q^2/2$. Under this 
circumstance where small transverse distances are being probed, 
the quark-pomeron interaction can no longer be considered as pointlike
and the strength $\beta_0$ must be replaced by a 
decreasing interaction, $\beta_0 f(k^2)$, with $f(k^2)$ a 
pomeron form factor. The deep reason for this phenomenological 
introduction of a form factor is that the QCD interaction of
a color neutral dipole of size $r$ vanishes like $r^2$ at small $r$. 
This colour transparency phenomenon implies that, at small transverse 
distances, quark additivity cannot hold in QCD. This aspect needs 
not be present in hadron-hadron soft interaction where the 
off-shellness of quarks, $k^2$, is limited by hadronic wavefunctions
and therefore $f(k^2)$ may always stay close to 1. 
  
We thus get for the photon-pomeron-rho vertex \[
2\beta_0f(k^2){ef_{\rho}M_{\rho}\over 2k^2}.
\]
The factor $k^2$ in the denominator comes from the quark propagator in 
Fig.~\ref{rho-prod}. Within the non-relativistic description of the 
$\rho$-meson, the remaining part of the quark loop in Fig.~\ref{rho-prod} 
is simply related to the $\rho$ decay constant, $f_{\rho}$, defined as 
\begin{equation}
\label{decay-constant}
\langle 0|J^{\mu}(0)|\rho(q,h)\rangle=
ef_{\rho}M_{\rho}\varepsilon^{\mu}(q,h),
\end{equation}
i.e. to the $\rho$ wave function at the origin. 

The differential cross section is obtained by putting all the factors 
together and squaring. With these inputs, the model provides a good 
description for the $Q^2$-behavior of the production cross section 
observed by EMC and NMC for the $\rho$-meson~\cite{don87}. From 
Eq.~(\ref{pomeron}), an energy dependence $\sim W^{0.32}$ is expected. 
This is at variance with the observed rising at HERA. I will return to 
this point in section 5.

\section{Towards a microscopic description of the pomeron}

Soon after the formulation of QCD, Low and Nussinov have proposed 
that pomeron exchange corresponds to soft gluon exchanges~\cite{low75}. 
The problem is then to treat the gluon dynamics in the nonperturbative 
domain of QCD. Here I consider the stochastic vacuum model~\cite{dos87}. 
This model and its application to high-energy scattering has been 
reviewed during the First ELFE Summer School~\cite{nac95}.

The model accomodates the non-trivial structure of the QCD vaccuum. 
Evidence for a non-trivial structure (as compared to a perturbative 
vacuum) is given by the sizeable value of the gluon 
condensate \[
\langle g^2FF\rangle=
\langle 0|{g^2\over 4\pi^2}F^a_{\mu\nu}(x)F^{a\mu\nu}(x)|0\rangle
\approx (0.4\,{\rm GeV})^4,
\] when compared to the QCD scale $\Lambda\approx 0.2\,$GeV. 
Within the model, the QCD vacuum is considered as a superposition 
of configurations where the gluon field strength condenses into 
domains (analogously to the magnetic field in a ferromagnet). 

The basic ingredient in the model is a gluon field strength 
two-point correlator. In order to define a gauge invariant 
correlator, one has to introduce the modified gluon field 
strength $F_{\mu\nu}(x,\omega)$ which is obtained from the field 
strength at point $x$ by parallel transporting the colour content 
to a reference point $\omega$
$$
F_{\mu\nu}(x,\omega)=\Phi^{-1}(x,\omega)F_{\mu\nu}(x)\Phi(x,\omega),
$$
with $\Phi(x,\omega)=P\exp[-ig\int_\omega^x Adz]$. Assuming that 
the main features of the correlator
$\langle0|F_{\mu\nu}(x,\omega)F_{\rho\sigma}(y,\omega)|0\rangle$ do not 
depend crucially on the choice of the reference point $\omega$ we 
obtain for the dependence on $z=x-y$ the most general form 
\begin{eqnarray}\label{correlator}
&&\langle g^2F^c_{\mu\nu}(x,\omega)F^d_{\rho\sigma}(y,\omega)\rangle
={\delta^{cd}\over N_c^2-1}{\langle g^2FF\rangle\over12}\Big\{\kappa
(\eta_{\mu\rho}\eta_{\nu\sigma}-\eta_{\mu\sigma}\eta_{\nu\rho})D(z^2/a^2)
\nonumber\\
&&+(1-\kappa){1\over2}\Big[\partial_{\mu}(z_{\rho}\eta_{\nu\sigma}
-z_{\sigma}\eta_{\nu\rho})+\partial_{\nu}(z_{\sigma}\eta_{\mu\rho}
-z_{\rho}\eta_{\mu\sigma})\Big]D_1(z^2/a^2)\Big\}.
\end{eqnarray}
The correlator incorporates the gluon condensate at $x=y$ and the 
fall-off of the correlation of two field strengths at large distances. 
This fall-off occurs on a typical correlation length, $a$, the typical 
size of the domain, which turns out to be about $0.3\,$fm.
The correlator $D$ is specific for a non-Abelian gauge theory 
since the homogeneous Maxwell equations
$$
\epsilon^{\mu\nu\rho\sigma}\partial_\nu F_{\rho\sigma}=0
$$
allow only the tensor structure associated with $D_1$ in 
Eq.~(\ref{correlator}), i.e. $\kappa=0$ in an Abelian theory without 
magnetic monopoles. 

The model leads to linear confinement, i.e. the Wison loop satisfies 
an area law for large contours, only if $\kappa\ne 0$. In this approach, 
confinement is then explicitely due to the non-Abelian nature of QCD. 
Notice that lattice simulations support the non-Abelian character of 
the correlator Eq.~(\ref{correlator}) by giving 
$\kappa\approx 0.74$~\cite{dig92}.

The high-energy scattering of two color singlet dipoles  $q_1\bar{q}_1$
and $q_2\bar{q}_2$ can be treated analogously to the situation of 
heavy quarks encountered in the Wilson area law. In the center of mass 
frame, each quark (or antiquark) travels with nearly the speed of light 
along some $z$-direction for $q_1\bar{q}_1$ or the opposite for 
$q_2\bar{q}_2$. During their travel, quarks interact with the vacuum 
field. This interaction shows up as a non abelian phase~\cite{nac91} \[
V=P\exp[-ig\int_\Gamma A dz]
\]where $\Gamma$ is the classical path of the quark (an antiquark gets 
a similar phase but its path is backward oriented). For dipole 1, the 
quark phase, $V_1$, together with the antiquark phase, $V_{\bar{1}}$, 
and the two gauge links entering the definition of the initial and 
final state dipole give rise to a Wilson loop $W_1$. The resulting 
dipole-dipole interaction is then the expectation value of the 
product of the two loops $W_1$ and $W_2$. 

\begin{figure}[h]
$$
\epsfxsize=69.5mm\epsfbox[140 100 700 480]{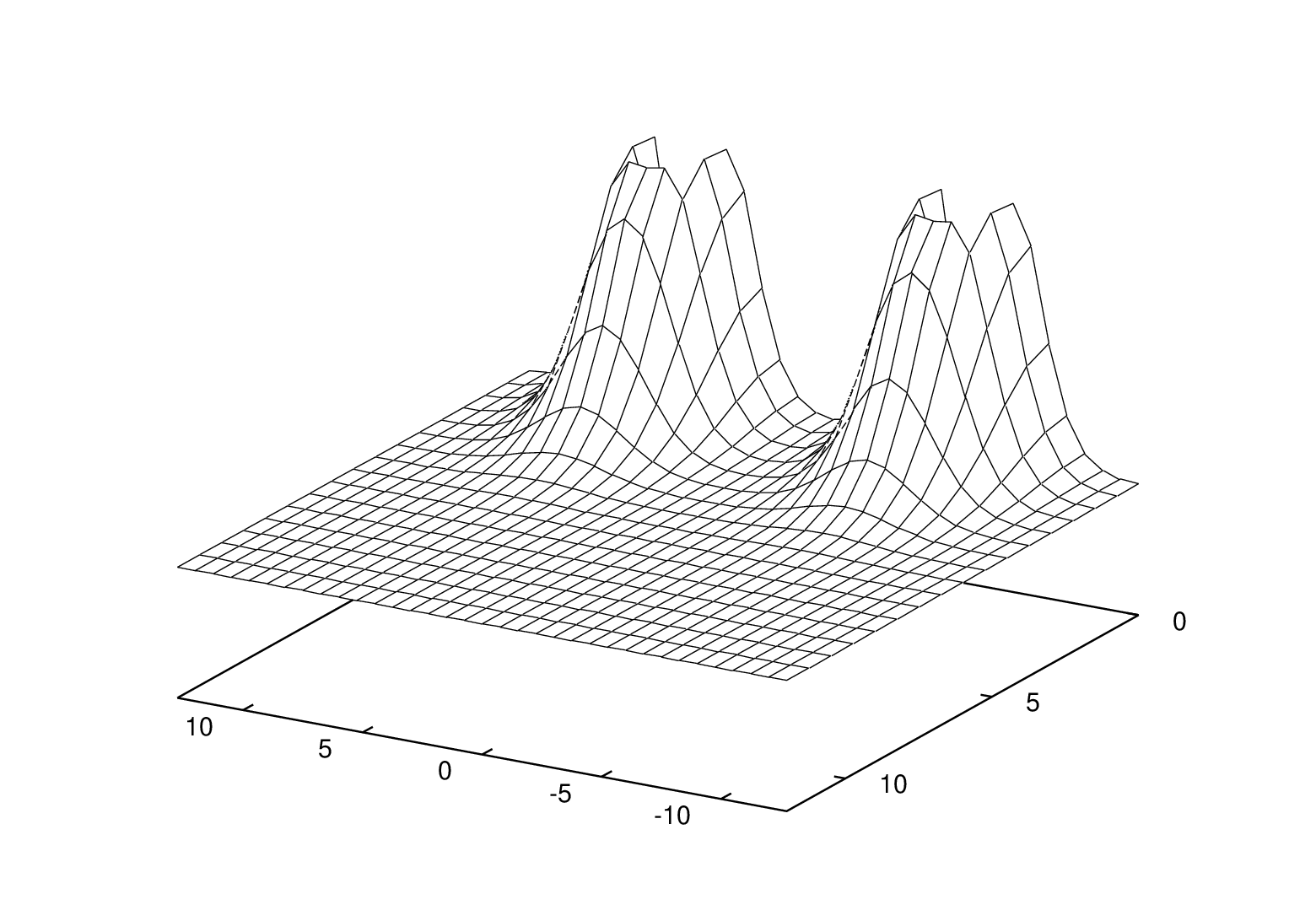}
\unitlength=6.95mm
\begin{picture}(0,0)
\put(-2,.5){\small$y/a$}
\put(-9.5,.5){\small$x/a$}
\end{picture}
\epsfxsize=69.5mm\epsfbox[140 100 700 480]{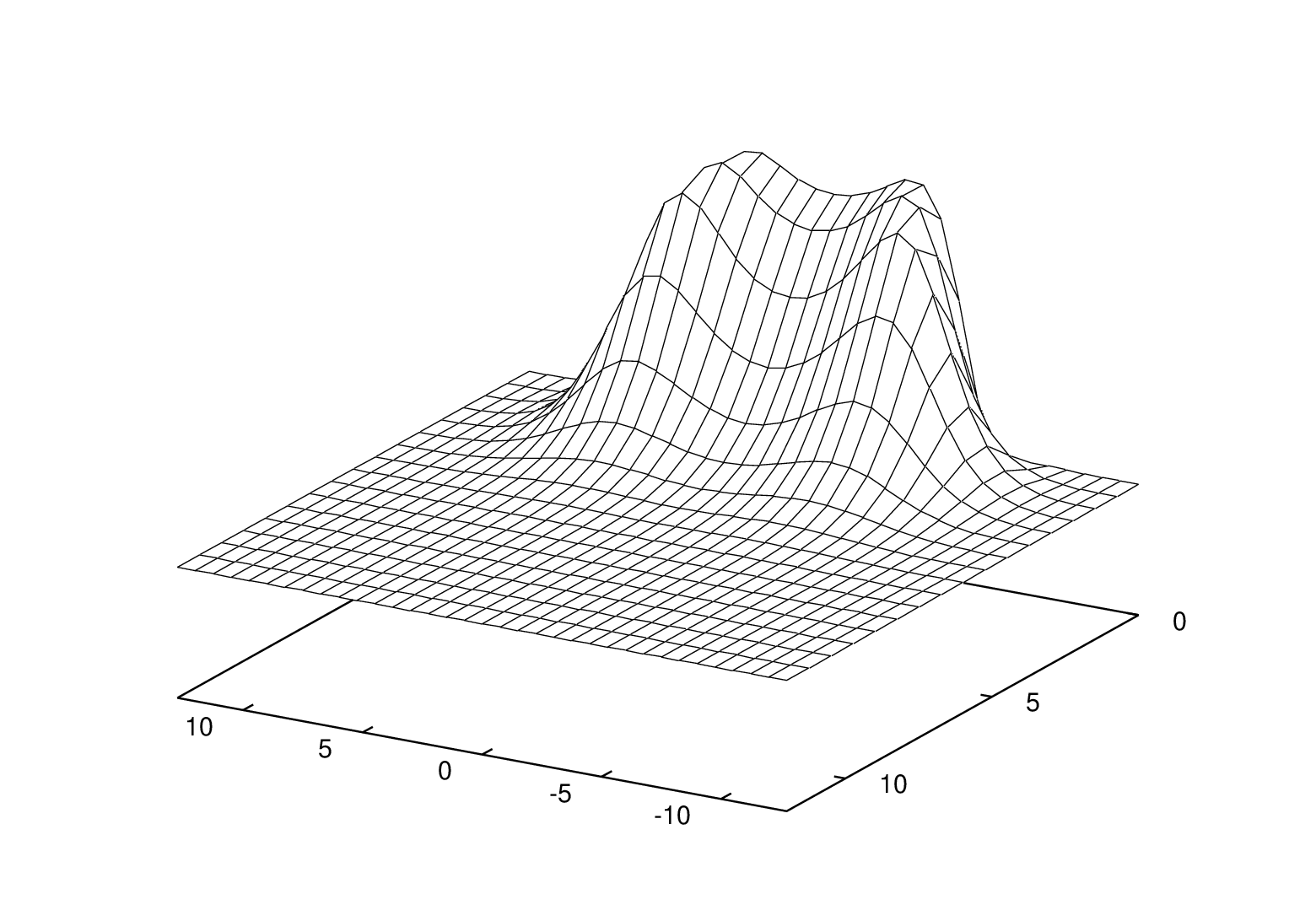}
\unitlength=6.95mm
\begin{picture}(0,0)
\put(-2,.5){\small$y/a$}
\put(-9.5,.5){\small$x/a$}
\end{picture}
$$
\caption{(a) Color interaction amplitude for $\kappa=0$ as a function 
of the impact position between the two dipoles. Dipole 1 has a transverse 
size $r_1=a$ and an average over its orientation is understood. Dipole 2 
has a transverse size $r_2=12\,a$ and lies along the $x$-axis. 
\protect\\
(b) Color interaction amplitude for $\kappa=1$.}
\label{color-interaction}
\end{figure}

In the model of the stochastic vacuum, this expectation value is 
related to the two-point correlator~\cite{dos94}. The two components 
of the correlator in Eq.~(\ref{correlator}) gives rise to distinct 
types of interaction. This is illustrated in Fig.~\ref{color-interaction} 
where the interaction amplitude between a large dipole target oriented 
along a given $x$-axis and a small dipole probe is plotted as 
a function of the impact position for both (a) the ``non-confining'' 
case, $\kappa=0$, and (b) the ``purely confining'' case, $\kappa=1$. For 
simplicity a sum over the orientation of the probe has been carried out.

The non-confining term gives rise to a quark-quark interaction, where 
color charges only interact if their trajectories enter in a common 
domain of the vacuum structure. The confining term gives a string-string 
interaction where the color dipoles (connected by their strings) 
interact as whole objects rather than as isolated endpoints. We 
notice that quark additivity arises from the non-confining part 
of the interaction provided hadrons are much bigger than the domain 
size, but that it does not if the confining term is important.

From the nonperturbative scattering amplitude of color dipoles with 
fixed lengths $r_1$ and $r_2$, a valence quark picture can be constructed 
by distributing the positions of the end-points of the strings 
according to hadronic wave functions~\cite{dos94}. Integrating out 
the nucleon constituent degrees of freedom leads to our model 
amplitude for the dipole-proton interaction amplitude to be inserted in 
Eq.~(\ref{amplitude}). 

\section{Comparison with experiment}

From the dipole-proton interaction amplitude just discussed and with 
the perturbative photon wavefunction and a model wave function to 
describe the vector meson, one can compute the differential 
cross section for either a longitudinal or a transverse polarization 
of the photon~\cite{dos96}. The discussion of our results for these 
quantities gives me the opportunity to show the present experimental 
status at NMC energies, i.e. $W\approx 10-20\,$GeV.

\begin{figure}[h]
$$
\epsfxsize=13.9cm\epsfbox[110 440 510 697]{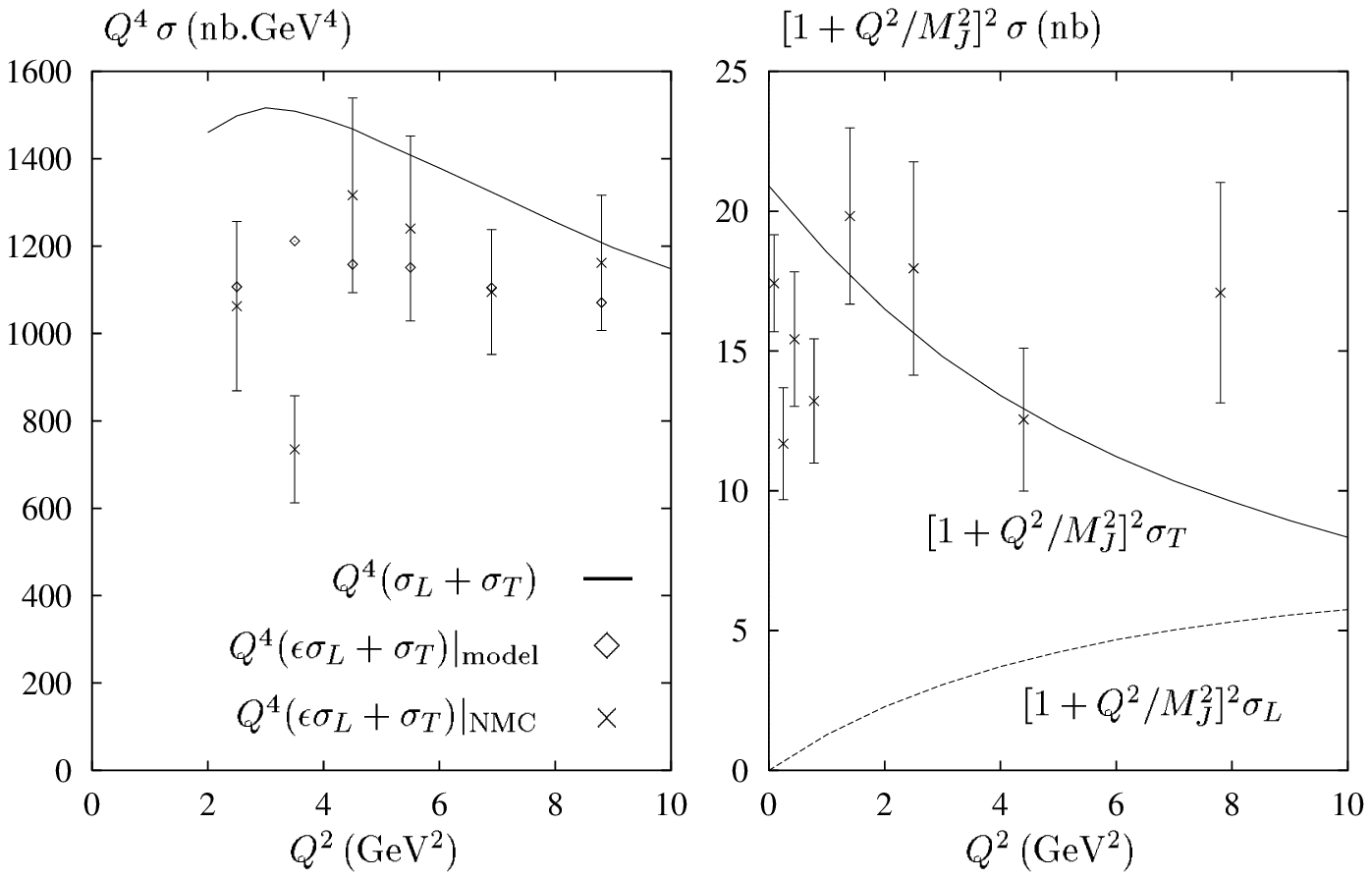}
$$
\caption{(a) The scaled cross section $Q^4\,\sigma(Q^2)$ for 
$\rho$-production in nb.GeV$^4$. The crosses are the 
NMC-results~\protect\cite{nmcrho} and the diamonds represent our 
prediction for the quantity $Q^4\,(\epsilon\sigma_L+\sigma_T)$ with 
the experimental polarization rate of NMC.\protect\\
(b) $J/\psi$-production cross section for longitudinal and 
transverse polarizations. To compare with EMC-data~\protect\cite{emcpsi}, 
one has to combine these two cross sections into 
$\sigma=\epsilon\sigma_L+\sigma_T$ with the polarization rate measured 
by EMC, $\epsilon\approx 0.7$.}
\label{integrated-cross-section}
\end{figure}

Let us first compare with the integrated cross section for $\rho$ 
and $J/\psi$-production. The experimental results come in the combination 
$\sigma_{\rm exp}=\epsilon\sigma_L+\sigma_T$, where $\epsilon$ is 
the rate of longitudinally polarized photons. The results as a function 
of $Q^2$ are show in Fig.~\ref{integrated-cross-section}. 

The model parameters, $\langle g^2FF\rangle$ and $a$, were adjusted 
to fit the proton-proton total cross section and elastic slope. Keeping 
them fixed, the magnitude of our cross sections depends on the meson 
wave function parameters. Once the value of the wave function at the 
origin is fixed to fit the meson decay constant, see 
Eq.~(\ref{decay-constant}), the most important ingredient is the 
transverse extension parameter for light quark systems and the quark 
mass when heavy quarks are involved. With the shrinkage of the photon, 
this influence decreases with increasing $Q^2$. The exact 
$z$-dependence of the wave function also affects the overall magnitude 
of the cross section.     

\begin{figure}[h]
$$
\epsfxsize=13.9cm\epsfbox[110 440 510 697]{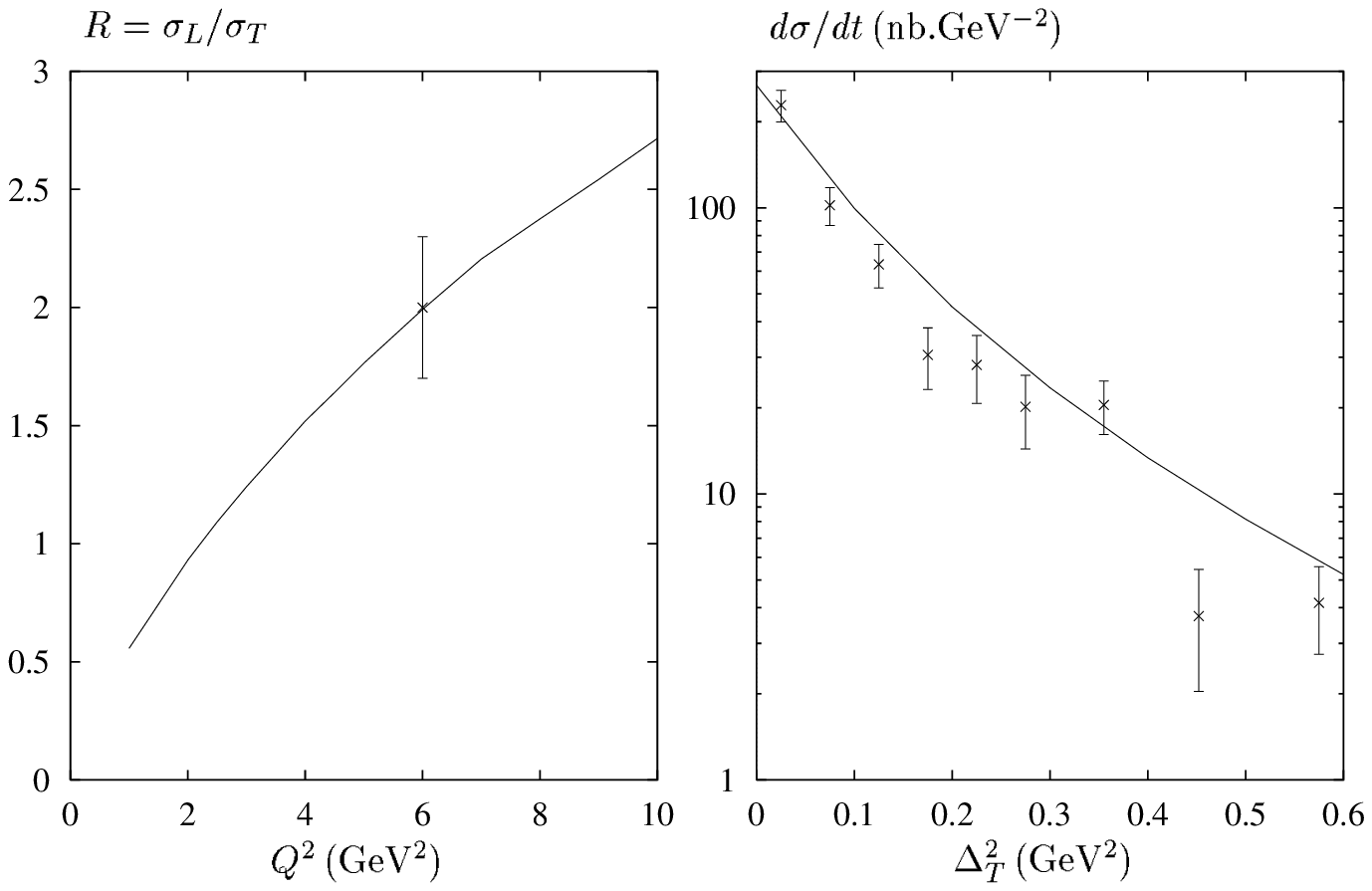}
$$
\caption{(a) The ratio for longitudinal to transverse cross section for 
$\rho$-production. The data point is from Ref.~\protect\cite{nmcrho}. 
Other data compare well within errors but they either are far outside
the 10--$20\,$GeV range or have large errorbars.\protect\\
(b) The differential cross section, $d\sigma/dt(\Delta_T^2)$, for 
$\gamma^*+p\to\rho+p$ at $Q^2=6\,$GeV$^2$. Data are from 
Ref.~\protect\cite{nmcrho}.}
\label{ratio-and-t-dependence}
\end{figure}

The $Q^2$-behavior of specific observables, such as the ratio of 
longitudinal to transverse cross sections, $R(Q^2)=\sigma_L/\sigma_T$, 
offers possibilities to distinguish between different models. The ratio 
is plotted in Fig.~\ref{ratio-and-t-dependence}(a). For large $Q^2$, 
$R=O(Q^2)$ but this behavior is not yet reached in the intermediate 
range where $R$ grows slower than $Q^2$.

Another important check is provided by looking at the $t$-dependence 
of the differential cross section. We show our result for 
$\epsilon d\sigma_L/dt+d\sigma_T/dt$ versus $\Delta_T^2$ at 6$\,$GeV$^2$ 
in Fig.~\ref{ratio-and-t-dependence}(b) and compare it to the NMC points 
for the deuteron outside of the coherent production region~\cite{nmcrho}. 
Notice also that there is a non trivial $Q^2$-dependence of the slope 
$B=d\ln(d\sigma/dt)/dt$ which comes with the slowly shrinking transverse 
size of the photon as $Q^2$ grows.

\section{Energy dependence}

\begin{figure}[h]
$$
\epsfxsize10cm\epsfbox[160 445 445 705]{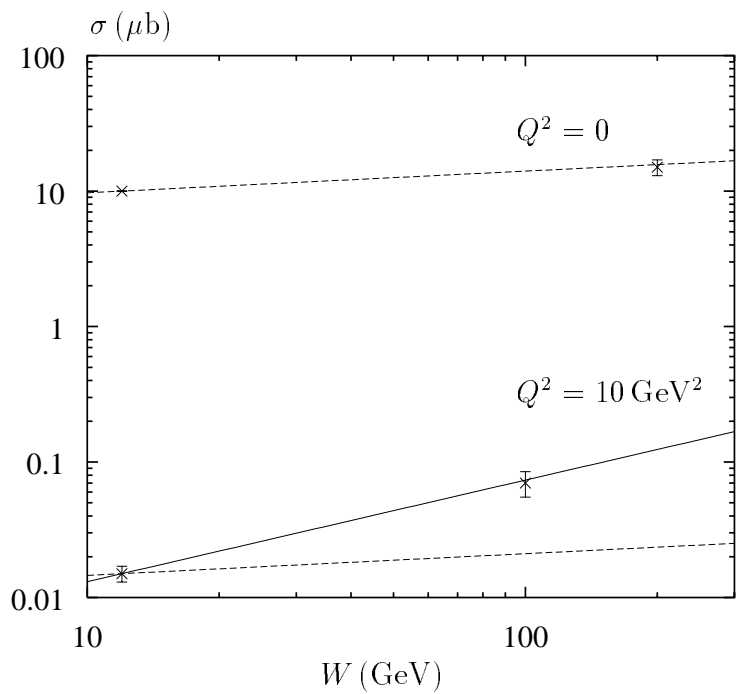}
$$
\caption{Photoproduction and electroproduction at $Q^2=10\,$GeV$^2$ of 
$\rho$-mesons. The energy rise observed in photoproduction is compatible 
with the soft pomeron phenomenology. A quite different behavior is 
observed at large $Q^2$}\label{hera}
\end{figure}

The pomeron phenomenology extracted from hadron-hadron scattering 
as in section 2 predicts for the exclusive vector meson electroproduction 
cross section a universal energy dependence $\sim W^{0.32}$ for large $W$. 
According to recent HERA results, the energy dependence at large $Q^2$ 
is however much more pronounced than such a ``soft'' behavior. The 
observed trend of the data is shown in Fig.~\ref{hera}.

Such a behavior has to be compared with the energy-dependence observed 
for the total photon-proton cross section, the latter also showing a 
weak dependence, $W^{0.16}$, for $Q^2=0$, and a strong one at large 
$Q^2$. The physics driving the rise of $F_2(x,Q^2)$, which is related 
to the total cross section through \[
F_2(x,Q^2)={Q^2\over\pi e^2}(\sigma_T+\sigma_L),
\] towards small $x_B\approx Q^2/W^2$ (i.e., at fixed $Q^2$, towards 
large energy) may then explain what occurs in 
electroproduction.\footnote{A closer look to transverse distances 
involved shows that exclusive vector meson production compares better 
to the longitudinal cross section $\sigma_L$ than to the sum 
$\sigma_T+\sigma_L$.} In the same spirit, it should be noted 
that the energy dependence in $J/\psi$-photoproduction is similar 
to large $Q^2$ $\rho$-production. A possible explanation for this 
similarity is that the typical transverse distances involved in both 
cases are the same. This can be deduced from the expression of the 
``photon size'' defined in section~1, 
$\varepsilon=[z(1-z)Q^2+m_q^2]^{1/2}$. In 
the $J/\psi$-case and at $Q^2=0$, the presence of the large $c$-quark 
mass leads to a typical transverse size probed as small as in the 
light quark case at some large $Q^2$ (a quantitative estimate gives 
$Q^2\approx M_{J/\psi}^2\approx 10\,$GeV$^2$). A study of the energy 
dependence in vector meson electroproduction in which the above 
connections appear naturally is given in Ref.~\cite{nem96}.

One can estimate the $t\to0$ limit of the amplitude
${\cal M}(q\bar{q}+p\to q\bar{q}+p)$ appearing in Eq.~(\ref{amplitude}) 
for small transverse distances. It is related, through the optical 
theorem\footnote{the real part of ${\cal M}$ can be neglected} \[
{\cal M}(s,r,t=0)=is\sigma^{\rm tot}(s,r),
\] to the total cross section of a small $q\bar{q}$-dipole, with 
transverse extension $r$, scattering on a proton. This in 
turn can be evaluated \[
\sigma^{\rm tot}(s,r)={\pi^2\over3}\alpha_S xG_p r^2,
\] where $G_p$ is the gluon density in the proton. 
Replacing this in Eq.~(\ref{amplitude}), one gets for the dominant 
part at large $Q^2$~\cite{rys93,bro94} \[
{\cal M}(s,t=0)=is{\pi^2\over3}\alpha_S xG_p
{\displaystyle 4ef_V\int_0^1 dz {\varphi_V(z)\over z(1-z)}
\over\displaystyle Q^3\int_0^1dz \varphi_V(z)}.
\]
Let us first notice that, the leading transition at large $Q^2\gg M_V^2$ 
is that of a longitudinal photon turning into a longitudinal vector 
meson. Other transition amplitudes are suppressed by some power 
of $1/Q$ and are asymptotically negligible in the cross section. 

In the limit of large $Q^2$, the meson wave function is probed 
only at 0-transverse distance, i.e. it is the distribution 
amplitude, $\varphi_V(z)=\psi_V(z,r_T=0)$, of the vector meson 
which matters in the meson wave function together with the meson 
decay constant $f_V$. Actually, it is only the quantity \[
\eta_V=\int_0^1 dz {\varphi(z)\over 2z(1-z)},
\] for a normalized $\varphi$, $\int_0^1 dz\varphi(z)=1$, which 
enters in the expression of the scattering amplitude. 
The distribution amplitude also occurs in exclusive reactions 
at large momentum transfer~\cite{bro89}, but in general 
inside different integrals. For the purpose of extracting 
distribution amplitude from experiments, the present 
reactions turn out to be complementary to hard exclusive ones.

One of the difficulties in the above derivation is to find how are 
the arguments $x$ and $\mu^2$ entering in the gluon density 
$xG_p(x,\mu^2)$ related to the physical $x_B$ and $Q^2$. One of the 
problem is that the above amplitude is explicitely computed at 
$\Delta=0$, a limit which turns out to be unphysical since the vector 
meson and the photon do not have the same mass. This difficulty can be 
circumvented by introducing off-forward parton densities~\cite{rad96}.

The remaining $t$-dependence of the amplitude has a soft origin 
and is not accessible via a perturbative treatment. Its description 
requires a non-perturbative approach such as the one presented 
in section 3.

\section{ELFE regime}

The experimental accuracy reached so far has revealed basic 
facts about the physics at work in vector meson production at 
small momentum transfer. However quantitative tests which 
are necessary to disentangle the various aspects of the reaction 
seem only feasible with high-luminosity facilities. 

The various approaches described above rely on high-energy approximation. 
At ELFE, the maximum center of mass energy is $W\approx 5\,$GeV, a 
value which lies just at the low end of our high-energy 
domain. From the Regge phenomenology point of view, exchanges 
of reggeons other than the pomeron are important in this energy 
domain. This aspect makes the microscopic description of the 
process much more difficult. To avoid this complication, it is 
interesting to focus on $\phi$ and $J/\psi$-production because 
the coupling of these particles to non-vacuum trajectories is 
strongly suppressed.

With the limitation $x_B\le 0.1$, the range of $Q^2$ is restricted 
to $Q^2\le 3\,$GeV$^2$, that is with ELFE one just reaches the 
``hard'' region where a simple $q\bar{q}$ description of the 
photon may be sufficient to describe the process. An understanding 
of the small $Q^2$ region is necessary. The $Q^2\to 0$ limit is 
commonly described through the vector meson dominance of the 
photon and, within the Donnachie-Landshoff model, it gives a 
reasonable description of $\rho$ and $\phi$-production.

Finally two very interesting extensions have not been discussed 
here. These are, on the one hand, the study of the reaction for 
nuclear targets instead of the nucleon and, on the other hand, 
the transition from low to high momentum transfer.

\ack

I thank Rainer Jakob for his careful reading of the manuscript. 
Part of this review describes a work carried out with H.G. Dosch, 
G. Kulzinger and H.J. Pirner. This work was supported by the Federal 
Ministry of Education, Science, Research and Technology (BMBF) under 
grant no. 06 HD 742.

\end{document}